\PassOptionsToPackage{super,sort&compress}{natbib}
\documentclass[twocolumn,prl,prl,amsmath,amssymb,reprint,superscriptaddress]{revtex4-1}
\usepackage{float}
\usepackage{color}
\usepackage{amsmath}
\usepackage[super,sort&compress]{natbib}
\usepackage{color}
\usepackage{epstopdf}
\usepackage{hyperref}
\usepackage{braket}
\usepackage[utf8]{inputenc} 
\usepackage[T1]{fontenc}
\usepackage{multirow}
\usepackage{graphicx}
\usepackage{dcolumn}
\usepackage{bm}

\begin{document}
\preprint{AIP/123-QED}

\title{Matter-wave interference of a native polypeptide}

\newcommand{\VCQ}{University of Vienna, Faculty of Physics, VCQ, Boltzmanngasse 5, A-1090 Vienna, Austria}
\newcommand{\IC}{Centre for Cold Matter, Blackett Laboratory, Imperial College London, Prince Consort Road, London SW7 2AZ, United Kingdom}
\newcommand{\SF}{Department of Chemistry and the PULSE Institute, Stanford University, Stanford, California 94305, USA}
\newcommand{\SL}{SLAC National Accelerator Laboratory, Menlo Park, California 94025, USA}

\author{A. Shayeghi}
\affiliation{\VCQ}

\author{P. Rieser}
\affiliation{\VCQ}

\author{G. Richter}
\affiliation{\VCQ}

\author{U. Sezer}
\affiliation{\VCQ}

\author{J. H. Rodewald}
\affiliation{\IC}

\author{P. Geyer}
\affiliation{\VCQ}

\author{T. J. Martinez}
\affiliation{\SF}
\affiliation{\SL}

\author{M. Arndt}
\altaffiliation{Corresponding authors: armin.shayeghi@univie.ac.at,\\ markus.arndt@univie.ac.at}
\affiliation{\VCQ}

\date{\today}

\begin{abstract}
The de Broglie wave nature of matter is a paradigmatic example of fundamental quantum physics and enables precise measurements of  forces, fundamental constants and even material properties. However, even though matter-wave interferometry is nowadays routinely realized in many laboratories, this feat has remained an outstanding challenge for the vast class of native polypeptides, the building blocks of life, which are ubiquitous in biology but fragile and difficult to handle. Here, we demonstrate the quantum wave nature of gramicidin, a natural antibiotic composed of 15 amino acids. Femtosecond laser desorption of a thin biomolecular film with intensities up to 1~TW/cm$^2$ transfers these molecules into a cold noble gas jet. Even though the peptide's de Broglie wavelength is as tiny as 350~fm, the molecular coherence is delocalized over more than 20 times the molecular size in our all-optical time-domain Talbot-Lau interferometer. We compare the observed interference fringes for two different interference orders with a model that includes both a rigorous treatment of the peptide's quantum wave nature as well as a quantum chemical assessment of its optical properties to distinguish our result from classical predictions. The successful realization of quantum optics with this polypeptide as a prototypical biomolecule paves the way for quantum-assisted molecule metrology and in particular the optical spectroscopy of a large class of biologically relevant molecules.  
\end{abstract}

\pacs{Valid PACS appear here} 
                            
\maketitle


The wave-particle duality of massive matter has become an important aspect of modern physics. Atom interferometry~\cite{Cronin2009,CLAUSER1997121} enabled new tests from quantum physics \cite{Kovachy2015} to general relativity \cite{Schlippert2014,Asenbaum2017}, cosmology \cite{Hamilton2015} inertial sensing \cite{Geiger2011,Savoie2018} precision measurements of fundamental constants \cite{Parker2018} and forces \cite{Haslinger2017}. The de Broglie wave nature has been shown for large molecules, from fullerenes~\cite{Arndt1999} and molecular clusters~\cite{Haslinger2013} up to even high-mass particles~\cite{Eibenberger2013}. Such experiments probe the quantum-to-classical interface and can even be used as a unique tool to characterize neutral molecules in the gas phase~\cite{Gring2010,Eibenberger2014} with the potential for minimally invasive high-precision spectroscopy~\cite{Rodewald2016}.

However until today, quantum optics with massive native biomolecules has remained elusive in particular due to the challenges in forming stable and intense molecular beams which can be detected with high efficiency and selectivity. Measurements on neutral biomolecules in the gas phase will, however, become valuable as they are solvent-free and allow predicting and evaluating biomolecular electronic properties independent of any coupling matrix environments~\cite{Jarrold2000a}. Here, we present the first realization of matter-wave interferometry of \emph{gramicidin A1}, a linear antibiotic polypeptide composed of 15 amino acids with a mass $m$~=~1882~amu~=~3.13~$\times~10^{-24}$~kg, naturally produced by the soil bacterium \emph{Bacillus brevis}. Interference experiments with this biomolecular prototype brings us a step closer towards quantum experiments with living organisms~\cite{Clauser1997}.

\begin{figure*}[!t]
		\resizebox{1.8\columnwidth}{!}{\includegraphics{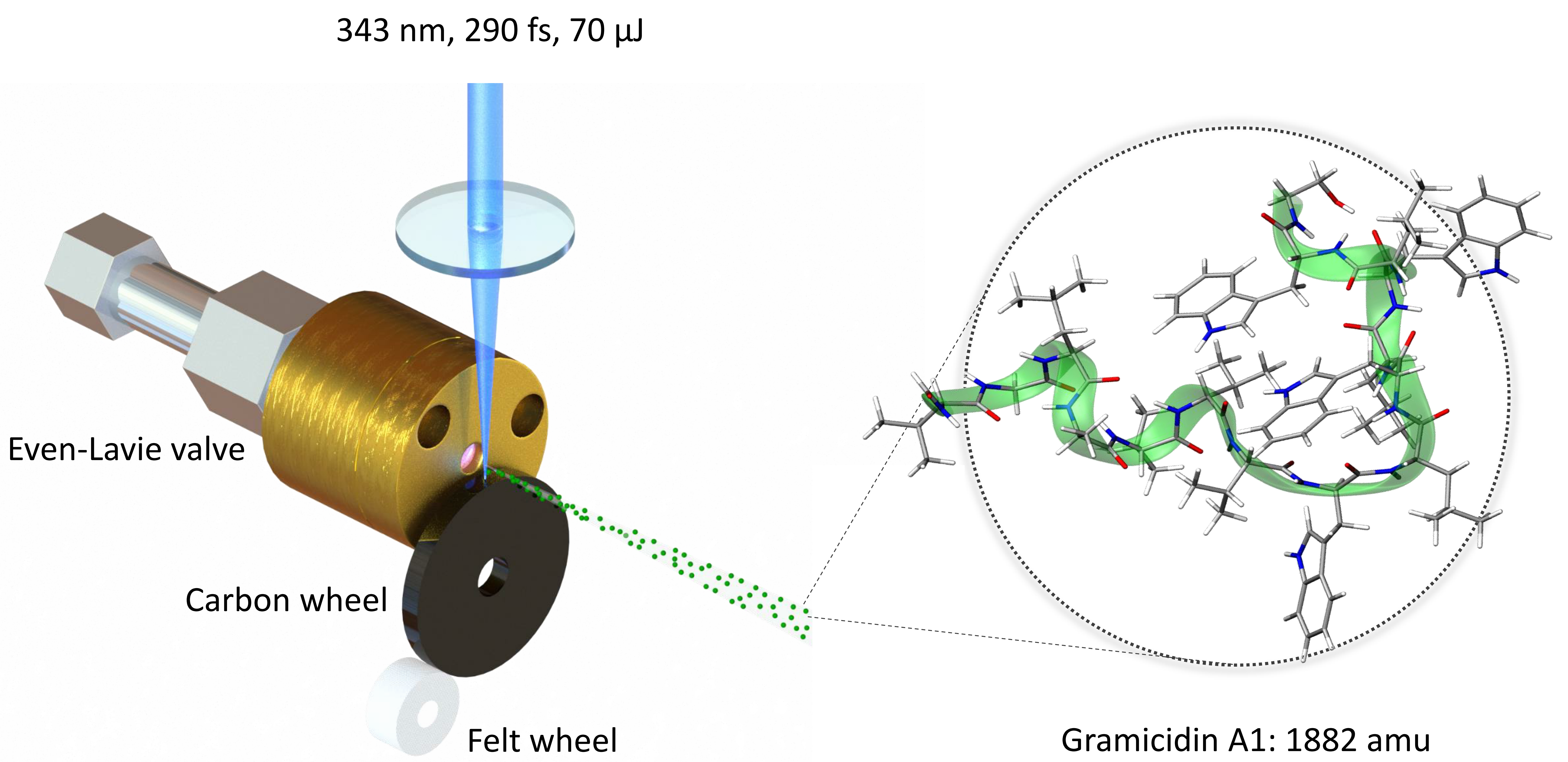}}
		\caption{Peptide source: Ultra-fast 293~fs laser pulses with an energy of up to 70~$\mu$J and a wavelength of 343~nm are focused to a spot diameter of 100~$\mu$m to desorb gramicidin molecules from a glassy carbon wheel. The molecules are picked up by an adiabatically expanding argon (helium) jet at 600~m/s (1200~m/s) from short-pulse high-pressure valve. The emerging polypeptide matter-wave has a de Broglie wavelength of 350~fm (175~fm). Gamicidin A1 is a 15 amino acid polypeptide. The green ribbon runs along the peptide bonds and the residues are shown as line diagrams. The four Tryptophan residues are the important chromophores that enable pulsed VUV laser ionization and thus the realization of optical diffraction gratings and photo-ionization in combination with mass-sensitive detection in our matter-wave interferometer.}
		\label{source} 
\end{figure*}

A typical matter-wave experiment requires an efficient source to launch neutral particles in high vacuum, beam splitters to coherently prepare, divide and recombine the quantum wave function associated with the molecular center-of-mass motion and an efficient detector with high sensitivity and mass resolution to record the result. 

For atom interferometry, these challenges have already been elegantly solved~\cite{Tino2014}. For interferometry with complex biomolecules, sources are a prime challenge. While evaporation and sublimation can still be used for vitamins~\cite{Mairhofer2017}, static thermal heating denatures and decomposes complex polypeptides. And while matrix assisted laser desorption (MALDI)~\cite{Karas1989} and electrospray ionization (ESI)~\cite{Fenn1989} can  volatilize even large proteins, they produce ions which are prone to  dephasing and decoherence in quantum experiments. Direct laser desorption using nanosecond laser pulses was proven useful to launch even neutral peptides into cold noble gas jets where selected species could be detected using photoionization by vacuum ultraviolet radiation~\cite{Marksteiner2009}.  
 
We use the idea presented in Figure~\ref{source}: A rotating felt wheel picks up a bimolecular powder and coats a glassy carbon wheel from which the molecules are desorbed by ultra-short pulsed laser light and entrained by a rapidly expanding noble gas jet. While nanosecond lasers were known to deliver intact peptide beams~\cite{Marksteiner2009}, we see a dramatic improvement in softness and efficiency by using ultra-fast laser pulses with TW/cm$^2$ intensities~\cite{Schaetti2018}.  The gramicidin beam is then skimmed, collimated and sent into the interferometer chamber with  $v=600$~m/s when using argon and $v=1200$~m/s when using helium. The different velocities are used to access different interference orders. 

Gramicidin thus arrives with a de Broglie wavelength of 350~fm (when using argon) which is about $10^4$ times smaller than the molecular van der Waals radius. We select a velocity spread of $\Delta v/v~\simeq~0.5\%$ and a longitudinal de Broglie coherence of ca. $200~\lambda_{\mathrm{dB}}\simeq 72~$pm. The initial transverse coherence is of the order of the thermal de Broglie wavelength, i.e. around 1~pm, too small for any practical matter-wave interference. However, transverse coherence can be prepared from initially incoherent particle ensembles using generalized Talbot-Lau interferometry. This idea is common in optics, was used in atom experiments~\cite{Clauser1994}, and has become the basis for molecule interferometry~\cite{Brezger2002,Gerlich2007,Gerlich2011}. 

Here we use an all-optical Talbot-Lau interferometer in the time domain (OTIMA)~\cite{Haslinger2013}, where nanosecond pulsed vacuum ultra-violet (VUV) standing light waves photo-deplete the molecular beam, if the molecules have ionization potentials $E_\mathrm{ion}<E_\mathrm{photon}$. Among all 20 natural amino acids, tryptophan is the only one that can be ionized by 7.9~eV light and the four tryptophan residues of gramicidin therefore allow efficient ionization of the peptide in our optical gratings and the detector~\cite{Schaetti2018}. 

\begin{figure*}[!t]
\centering
		\resizebox{2\columnwidth}{!}{\includegraphics{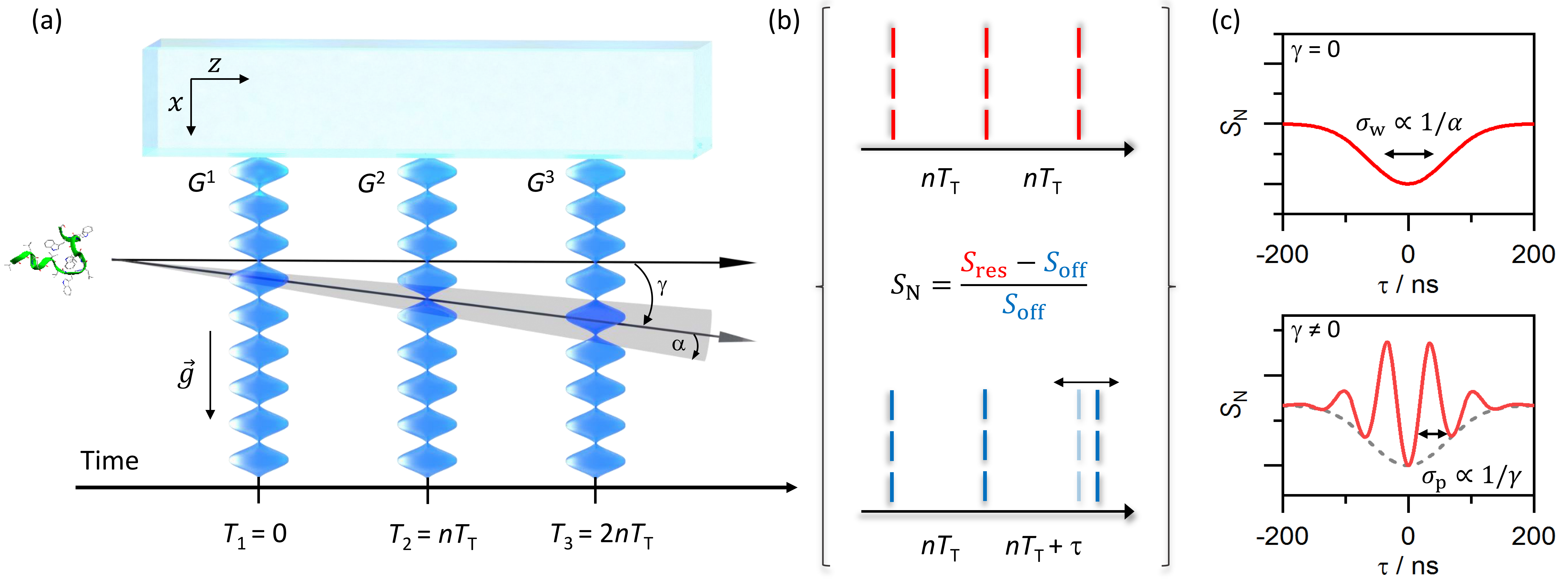}}
		\caption{Matter-wave interferometer (a): Three retro-reflected vacuum ultraviolet laser beams realize the standing light waves as well-defined pulsed photo-depletion gratings~\cite{Reiger2006}. The antinodes in $G^{(1)}$ prepare a comb of tightly confined  positions from where a molecule may emerge. Because of this projective confinement the wave coherence rapidly expands in free flight to cover several nodes and antinodes by the time the second grating fires. Rephasing of the matter-wave behind $G^{(2)}$ then leads to de Broglie interference of each molecule with itself and to the formation of a periodic molecular density pattern around the time when $G^{(3)}$ is fired. Only molecules whose wave functions are aligned with the nodes of $G^{(3)}$ are transmitted to the detector. The coherent rephasing occurs around a characteristic timescale, the $n$-th multiple of the Talbot time. A typical measurement (b): we toggle between two mass spectrometric modes : a symmetric mode (resonance), where pulse separation times are kept equal and close to $nT_{\mathrm{T}}$, and a asymmetric mode (off-resonant or reference), where we set an imbalance of some tens of nanoseconds. Imprinted fringes (c): In case the molecular beam velocity has a component parallel to $x$, the fringe pattern effectively has a transverse velocity component and its position relative to the third grating becomes time dependent. A fringe pattern is visible in case the divergence angle $\alpha$ is smaller than the tilt angle $\gamma$.}
		\label{tilt} 
\end{figure*}

The interferometer (Figure~\ref{tilt}) is composed of  three such  gratings $G^{(1)}-G^{(3)}$ formed by reflecting three  F$_2$ laser beams from the surface of a single dielectric mirror. Using a common mirror for all gratings ensures high vibrational stability and common mode rejection but it also impedes spatial scanning of the matter-wave fringes. However, one can still scan them in the time-domain~\cite{Haslinger2013,Rodewald2018}. For that purpose, we fix the pulse separation time $T$ between the gratings and use two complementary measurement settings. In the \textit{resonant} mode the pulse separation times $T=T_3-T_2=T_2-T_1$ are set to the $n$-th multiple of the Talbot time 
\begin{align}
nT_{\mathrm{T}}=~n\frac{md^2}{h},
\end{align}

with the grating period $d = \lambda_{\mathrm{L}}/2$ depending on the grating laser wavelength $\lambda_{\mathrm{L}}$ the particle mass $m$ and Planck's constant $h$. Interference then enhances or reduces the transmitted  molecular signal $S_{\mathrm{res}}$ depending on the  position of the matter-wave fringes relative to the antinodes of $G^{(3)}$. In the \textit{off-resonant} (reference) mode, the pulse separation time is kept imbalanced with $\Delta T\geq$~100~ns. For our molecular beam divergence and mass, this is  sufficient to smear out the interference pattern. The interference contrast is then defined as the normalized signal difference $S_{\mathrm{N}} = (S_{\mathrm{res}}-S_{\mathrm{off}})/S_{\mathrm{off}}$~\cite{Haslinger2013}. 

To model the expected signal we use a phase space description based on the Wigner function $w(x,p_x)$~\cite{Nimmrichter2011a} and modify it to account for tilted and divergent molecular beams (see Methods) including previous refinements modeling mirror and grating imperfections \cite{Doerre2015}. This allows comparing the experiment with both the quantum and the classical expectation in the same framework. Within this phase space formalism, the free evolution of a particle with initial position $x$ and  momentum $p_x$ is described by a shearing transformation $w\left(x,p_x\right)~\rightarrow~w\left(x-p_xt/m,~p_x\right)$. Additional terms may be introduced to account for earth's gravitational acceleration $g$ or a tilt of the molecular beam by an angle $\gamma$ with respect to the mirror surface, which results in a constant transverse momentum $p_{\gamma}=mv\tan(\gamma)$:
\begin{equation}
w\left(x,p_x\right) \rightarrow w\left(x-\frac{p_xt}{m}+\frac{p_{\gamma}t}{m}+\frac{gt^2}{2},p_x-p_{\gamma}-gmt\right).
\label{free_propagation_a_v}
\end{equation}

One can treat the motion along $z$ as a classical parameter and constrain the quantum analysis to a one-dimensional problem in which we concatenate the molecule's interaction with the first grating $G^{(1)}$ -- described by a transmission function $t^{(1)}$ -- its free evolution to $G^{(2)}$, the transmission $t^{(2)}$ through the second grating and the free evolution to $G^{(3)}$. The molecular density pattern at the third grating is the integral of $w_3(x,p_x)$ over $p_x$. Convolving this with the transmission of $G^{(3)}$ yields the signal seen by the mass spectrometer 
\begin{align}
S\left(\Delta x\right)=\sum_lS_{l}~\exp\left[\frac{2\pi il}{d}\Delta x(\Delta x_{\text{s}},T,\tau)\right].
\label{Signal}
\end{align}

It is periodic in $d$, depends on the Talbot time $T_{\mathrm{T}}$, the pulse delay $\tau$  of the grating with respect to the Talbot time  and the relative grating shift $\Delta x_s$. To probe this  pattern  without shifting any grating  we exploit the specific nature of  time-domain interferometry and tune $\tau$, i.e. the relative timing of $G^{(3)}$ around its resonant symmetric position~\cite{Rodewald2018}.
A quantitative description of the detected signal is more involved, in particular when a divergent and tilted beam is modeled including grating and mirror imperfections (see Methods). One finds that the transmission must depend 
on the dimensionless parameter $\beta$~\cite{Nimmrichter2011a}, which describes the ratio of the molecule's wavelength-dependent absorption coefficient $\sigma(\lambda_\mathrm{L})$ and its optical polarizability $\alpha(\lambda_\mathrm{L})$:
\begin{align}
\beta =\frac{n_0^{(\text{k})}}{2\phi_0^{(\text{k})}}=\frac{\lambda_{\text{L}}}{8\pi^2}\frac{\sigma(\lambda_{\mathrm{L}})}{\alpha(\lambda_{\mathrm{L}})}.
\label{Beta}
\end{align}

The absorption cross section determines the ionization probability and thus controls the effective opening fraction in all three VUV gratings. The polarizability at 157.6~nm determines the phase gramicidin molecules acquire during their transit through $G^{(2)}$. This phase leaves the fringe periodicity unchanged but modulates its contrast. The difference between the quantum and the classical expectations is  encoded in how these optical properties enter the  transmission function (see Methods).

A thorough understanding of the final signal therefore requires knowledge about the electronic properties of gramicidin with respect to its ground and excited states. This is a challenge since gramicidin has many possible conformational states and one has to evaluate electronic properties for an ensemble populating a complex potential energy surface (PES). The gramicidin molecule contains 1010 electrons which renders electronic structure calculations difficult even without global optimization of the conformational space and when combined with density functional theory (DFT). Here, we perform short ab-initio molecular dynamics (AIMD) simulations to roughly scan the conformational PES and to get a sense of the range of the dynamic polarizability of gramicidin (see Methods).  
Molecular geometries are extracted from the AIMD simulation every picosecond and are fed into subsequent DFT calculations to estimate the ensemble average of the optical polarizability $\left\langle\alpha(\lambda_{\mathrm{L}})\right\rangle_\mathrm{300K}$. Also the absorption cross section has to be considered as a thermal average $\left\langle\sigma(\lambda_{\mathrm{L}})\right\rangle_\mathrm{300K}$ in the calculation of $\beta$. It is therefore measured under identical experimental conditions in an independent experiment by monitoring the gramicidin ion count rate $N_{\mathrm{I}} = N_{\mathrm{0}} (1-\exp({-\sigma_{\mathrm{PI}}}\phi))$ as a function of the VUV photon fluence $\phi$ (see Methods).

These tools at hand, we can now analyze the matter-wave interferogram obtained with gramicidin, both in the first ($n=1$) and fractional ($n=1/2$) Talbot order. We record them by delaying the last grating in the 'resonant' interference mode in steps of 20~ns ($n$~=~1) and 10~ns ($n$~=~1/2) at a fixed time delay for the off-resonant reference signal by $\tau_{\mathrm{off}}$~=~200~ns ($n$~=~1) and 100~ns ($n$~=~1/2). For a finite collimation and tilt of the molecular beam, the fringe density pattern scans across the grating when $\tau$ is varied. We therefore expect a sinusoidal modulation with a Gaussian envelope~\cite{Rodewald2018}
\begin{align}
					S_{\mathrm{N}} = V_0 \exp{\left[-\left(\frac{\tau}{\sigma_{\mathrm{w}}\sqrt{2}}\right)^2\right]}\cos\left(2\pi\frac{(\tau-\tau_{\mathrm{off}})}{\sigma_{\mathrm{p}}}\right).
					\label{fit}
\end{align}

  The modulation of the fringe visibility  $V_0$  with $\tau$ allows us to  extract the divergence angle $\alpha$~=~0.4~mrad from the width $\sigma_{\mathrm{w}}$ of the resonance dip
\begin{align}
					\alpha = \arcsin{\left(\frac{d}{2v\sigma_{\mathrm{w}}\sqrt{2\ln{10}}}\right)}
\end{align}

  and  the tilt angle $\gamma$~=~1.7~mrad from the observed fringe period $\sigma_{\mathrm{p}}$
\begin{align}
					\gamma = \arcsin{\left(\frac{d}{v\sigma_{\mathrm{p}}}\right)}.
\end{align}	

\begin{figure}[!t]
		\resizebox{1\columnwidth}{!}{\includegraphics{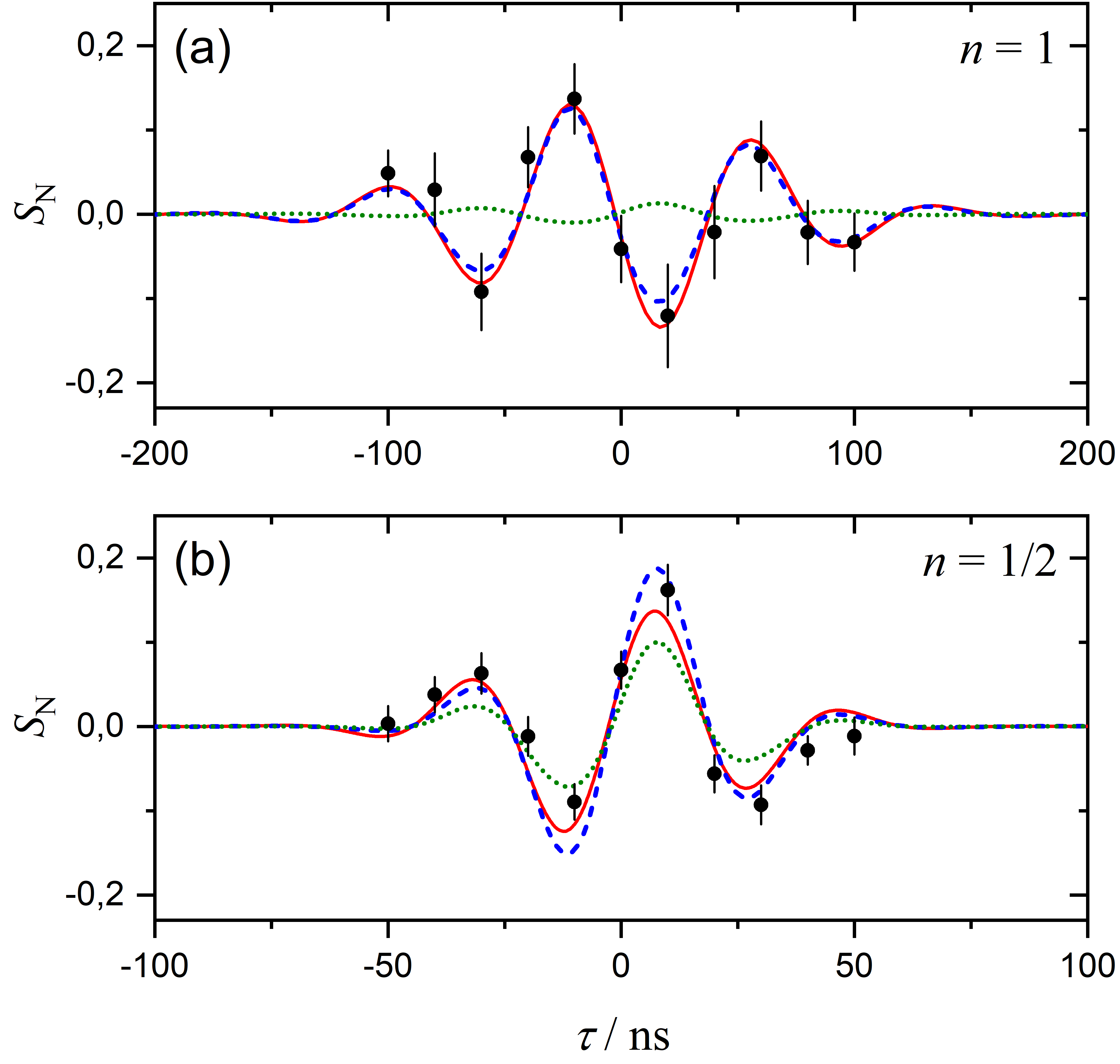}}
		\caption{Molecular interference patterns of gramicidin in the first (a) and half (b) Talbot order (black circles). Additionally, a fit according to (4) is shown (solid red line) together with a quantum- (dashed blue line) and a classical predicition (dotted green line). The fringes appear on the time-domain resonance dip when the pulse separation time between $G^{(2)}$ and $G^{(3)}$ is varied by a small delay $\tau$ around the Talbot resonance for the case of a tilted molecular beam. The envelope of the resoance dip is determined by the molecular divergence angle while the fringe period is determined by the tilt angle with respect to the mirror surface. Note the different scaling of the abscissa in (a) and (b).}
		\label{Dip} 
\end{figure}

We extract the model parameters $\alpha$ and $\gamma$ from the data in Figure~\ref{Dip}, the absorption cross section from independent measurments and the VUV polarizability from our quantum chemical analysis. The final result is shown in Figure~\ref{Dip}, with the experimental data points  (black circles), a fit based on (4) (solid red line), the quantum simulation (dashed blue line) and  the classical description (dotted green line). The experiment is very well described by quantum theory while the classical simulation fails to reproduce the data by a great margin, in particular close to the Talbot resonance. This quantum-classical discrepancy is expected to be less prominent at fractional Talbot orders, where the quantum fringe system may be approximated by a geometrical shadow. To address the $n$~=~1/2  order, argon is replaced  by helium which doubles the  mean velocity from 600 to 1200~m/s and allows observing interference fringes with half the pulse separation time at an unchanged interferometer length (see Methods).

The consistency of all data corroborates the hypothesis that the observation is due to genuine quantum interference.
And yet one may ask by what factor the assessment of the molecular parameters should be wrong to allow the data to be compatible with the classical model. We find that  $\beta\simeq 100$ would be required. This is two orders of magnitude off and appears unreasonable considering  $\beta$ of different organic molecules and clusters~\cite{Doerre2015}.

Interestingly, gravity plays an important role: The maximum normalized signal difference $S_{\mathrm{N}}$ on resonance is a function of the pulse separation time $T$. The effect of Earth's gravity on the fringe displacements  follows $S_{\mathrm{N}}(T) = V_0 \sin{\left(\frac{2\pi}{d}(b-gT^2)\right)}$ and is only significant for measuring $S_{\mathrm{N}}$ across several interference orders. This influence of gravity on the visibility at resonant timings has  been  used before to establish a proof-of-principle test of the weak equivalence principle for different isotopes of tetraphenylporphyrin~\cite{Rodewald2018}. However, the maximum visibility can be extracted in case of interference measurements on tilted molecular beams.  

In summary, we have demonstrated for the first time matter-wave interferometry with a complex native polypeptide, the antibiotic gramicidin. The fringe visibilities of around 20\% in both the first and the 1/2~th Talbot order stand in marked discrepancy to a classical phase space description but in very good agreement with quantum expectations based on our refined phase space model including a thorough quantum chemical analysis of the molecular electronic properties. This success is largely due to novel source techniques based on UV femtosecond desorption which can volatilize fragile biomolecules in a softer and more efficient way than other methods to date. The new source methods presented based on ultrafast, high-intensity desorption shall also pave a path to experiments with small proteins such as insulin in a similar setting. While matter-wave experiments with biomolecules in the gas phase do not elucidate biological function per se -- which is related to electronic structure determining folding dynamics and interactions with matrix environments -- our experiments show that quantum phenomena can be observed with such building blocks of life under suitable boundary conditions. Future experiments will extend this feat to proteins and DNA fragments, also to exploit the molecular interference pattern as a flying nanoruler for biomolecule metrology. The emerging field of quantum-assisted molecule metrology will become relevant for the study of optoelectronic properties of neutral biomolecular systems, that have not been accessible  by classical experiments to date.


\section{Acknowledgements}
A.S. acknowledges funding by the Austrian Science Fund (FWF) within the Lise-Meitner fellowship M~2364. M.A. has received funding from the European Research Council (ERC) under the European Union's Horizon 2020 research and innovation program (Grant Nr. 320694). U.S. was supported by the Austrian Science Fund (FWF)  within program W1210-N25. The computational results presented were partly obtained using the Vienna Scientific Cluster (VSC) within 70918. We are grateful to NSF for computing time provided on XSEDE resources via award TG-MCB090110. We are thankful to Nadine Asenbaum-D\"orre, Philipp Haslinger,  Andrea Grimaldi, and Gustavo G. Rondina for fruitful discussions and technical support. 

\section{Author contributions}
The experiments were conceived by A.S. and M.A. The interference experiments were realized by A.S., G.R. and P.R. The source has been designed and adapted to the OTIMA experiment by A.S., G.R., P.R., U.S. and P.G. Data analysis was performed by A.S., G.R., J.R., and  P.R. and  Quantum chemical simulations were modeled and analyzed by A.S. and T.M. and realized by T.M. The manuscript was written by A.S. and M.A. with contributions from all authors.


%

\end{thebibliography}

\section{Additional information}

Supplementary information is available for this paper at https://doi.org...
Correspondence should be addressed to A.S. and M.A.

\section{Methods}

\subparagraph{Sample preparation:}
Gramicidin D (Sigma Aldrich, CAS: 1405-97-6) is used which is a mixture of different antibiotic compounds. The major component is gramicidin A1, a linear polypeptide composed of 15 amino acids. It has the chemical formula C$_{99}$H$_{142}$N$_{20}$O$_{17}$. The molecule sketched in Figure 1 represents one specific configuration of gramicidin A1. The green ribbon follows the peptide sequence while the  tryptophan, valine and isoleucine rest groups are explicitly shown. The source emits a large variety of structural conformers, which all contribute to the same matter-wave interference pattern, since their mass and VUV optical properties are nearly identical. There are both fermions and bosons in the sample. All molecules can be excited in several of their  828 vibrational modes and highly excited in their rotational degrees of freedom,  However, indistinguishability between different molecules is irrelevant for genuine single-particle interference, that prevails essentially in all matter-wave experiments to date, even those with quantum degenerate gases.

\subparagraph{Molecular beam:}
The experiment runs at 100~Hz. In every cycle, an Even-Lavie valve releases a 20~$\mu$s short and dense pulse of argon with a backing pressure of about 30~bar. A femtosecond laser (Topag PHAROS, 290~fs, 70~$\mu$J, 343~nm) is focused ($\varnothing$~=~100~$\mu$m) on  the surface of a glassy carbon wheel coated with gramicidin which creates a plume of isolated molecules. The argon (helium) pulses then entrain the molecules with a mean velocity of around 600~m/s (1200~m/s). Further downstream the particle beam is skimmed (Beam Dynamics skimmer, $\varnothing$~=~2~mm), collimated to a rectangular shape of 0.6$\times$1~mm$^2$ (the longer axis parallel to the grating vectors) and finally transferred to the interferometer chamber via a differentially pumped stage. The pressure in the main chamber is 2$\times10^{-9}$~mbar, in the source chamber 1$\times10^{-8}$~mbar. Molecular beam velocities are determined by comparing the timing of the desorption laser with the detection laser pulse. 

\subparagraph{Grating lasers:}
The grating laser beams are emitted by three GAM EX50 fluorine lasers ($\lambda_{\mathrm{L}}$~=~157.6~nm, 4~mJ, 8~ns). All beams are reflected by the same dielectric mirror 3$\times$5~cm$^2$, coated onto a 2~cm thick CaF$_2$ substrate with the best technically available reflectivity in this wavelength range to date, i.e. $R\simeq 97\%$. The laser waists are elongated parallel to the molecular beam axis $z$ (10$\times$1~mm$^2$) and spatially separated by $\approx$2~cm which allows molecules of different velocities to interact with the same laser gratings at the same time. 

\subparagraph{Mirror imperfections:}
For a perfectly flat mirror and in the absence of external accelerations, $S_{\mathrm{N}}$ is positive and equal to the theoretical visibility. In a real world scenario, $S_{\mathrm{N}}$ is given by the visibility of the molecular density pattern at the position of $G^{(3)}$ and its relative displacement $\Delta D = \Delta x_1-2\Delta x_2+\Delta x_3$ of this Talbot image with respect to $G^{(3)}$. Here, $\Delta x_i$ captures both the possible mirror corrugations at either grating side or the displacement of the molecular fringe due to external accelerations -- for instance gravity. The molecular transmission is maximized for  $\Delta D = nd$ and minimized for $\Delta D = (n+1/2)d$ with $n \in \mathbb{Z}$. If the mirror surface had deformations exceeding 10~nm across the 10~mm grating laser beam profile, molecules of the same gas pulse but different velocities would experience differently shifted interferometers. To avoid the ensuing reduction in fringe contrast, the effective interaction region is set to $<2$~mm by the geometry of the final detection laser. 


\subparagraph{Coriolis force:}
Considering the current orientation of the molecular beam (N48,22$^\circ$ and particles fly SSE (162$^\circ$)) and assuming a velocity of 600~m/s at about 2.000~amu, the Coriolis force causes a shift of around 1~nm. Therefore, even for large velocity spreads phase averaging can be neglected. The Coriolis force also results in a path length difference which can be neglected too for the mass considered here. 


\subparagraph{Optical polarizability at $\lambda_{\mathrm{L}}$:}



The AIMD simulations are performed using the TeraChem  program package \cite{Terachem1, Terachem2}. During the AIMD run, a single molecule is propagated over 50~ps in time steps of 1~fs at a temperature of 300~K, which is controlled by a Bussi-Parinello thermostat \cite{Bussi2007c} with a relaxation time of 0.1~ps. The dynamic polarizability is computed by Q-Chem \cite{qchem} using DFT with the range-separated hybrid exchange-correlation functional LC-$\omega$PBEh \cite{Rohrdanz2009} and the 6-31G basis set. The Coupled-Perturbed Kohn-Sham method \cite{cpks} is used to calculate the optical polarizability for every extracted geometry at $\lambda_{\mathrm{L}}$ to obtain the ensemble average $\left\langle\alpha(\lambda_{\mathrm{L}})\right\rangle_\mathrm{300K}$ = $4\pi\epsilon_0$~(157~$\pm$~1)~\AA$^3$. 

\subparagraph{Ionization cross section at $\lambda_{\mathrm{L}}$:}

\begin{figure}[!t]
		\resizebox{1\columnwidth}{!}{\includegraphics{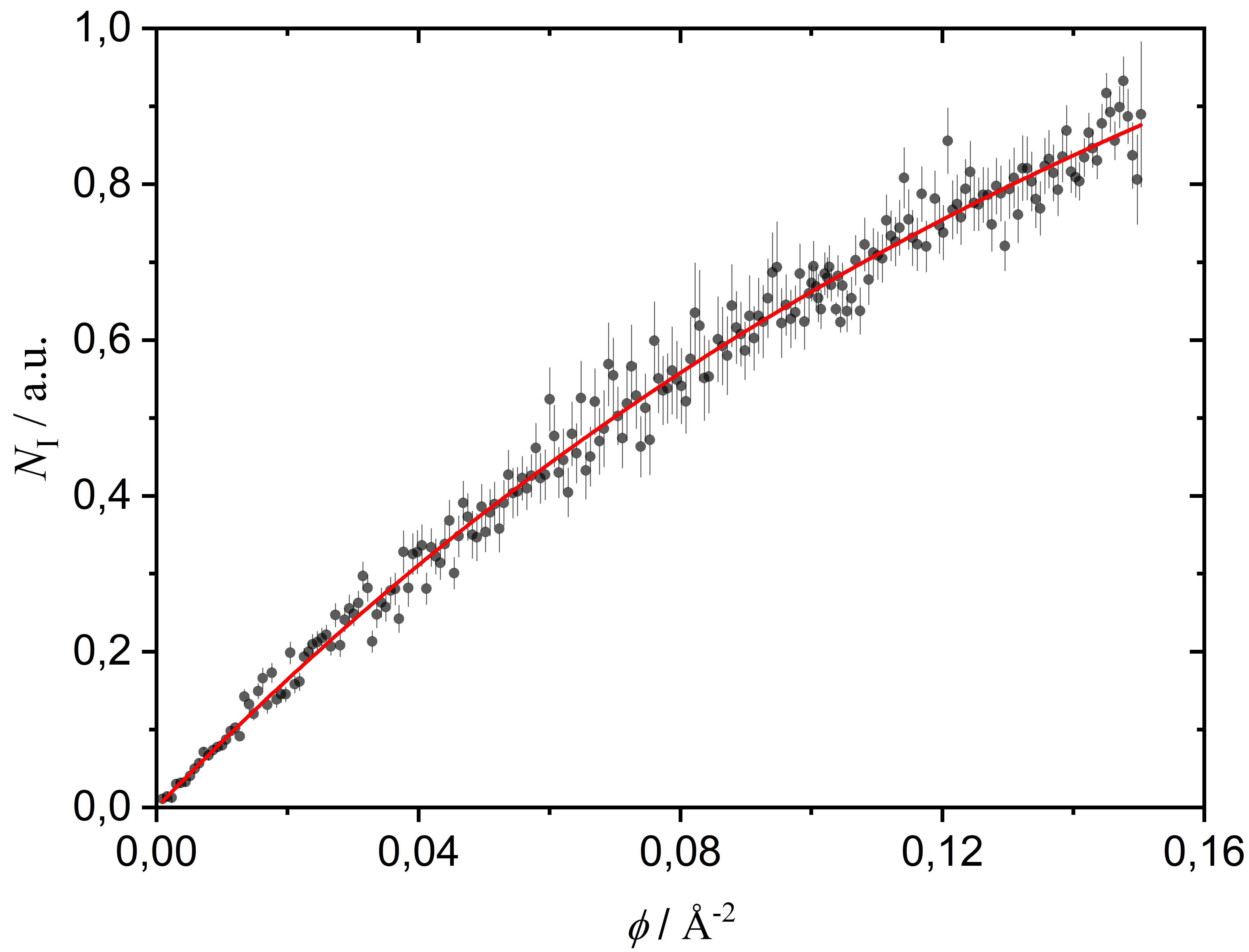}}
		\caption{Ion counts as a function of the photon fluence $\phi$. From a fit to the data points the ionization cross section $\sigma(\lambda_{\mathrm{L}})$ can be extracted according to Eq.~\ref{fit}. It can further be noted from the exponential  behaviour that in the presented range of available VUV intensities only single-photon processes occur.}
		\label{cross} 
\end{figure}

It should be noted that  the relevant relaxation channels  after absorption   are  ionization and  disscociation, since our detector is only sensitive to depletion of the molecular beam: $\sigma = \sigma_{\mathrm{PI}} + \sigma_{\mathrm{PD}}$, where $\sigma_{\mathrm{PI}}$ and $\sigma_{\mathrm{PD}}$ are the photoionization- and the photodissociation cross sections, respectively. We conclude that $\sigma_{\mathrm{PI}}$ is the main contribution since we did not detect any fragments upon irradiation with 157~nm light. However, $\sigma_{\mathrm{PI}}$ should still be considered as a lower limit to the total absorption cross section. In order to obtain $\sigma_{\mathrm{PI}}$, we measure and plot the number of counted ions $N_{\mathrm{I}}$ as a function of the photon fluence $\phi$ (see Figure \ref{cross}), which is the total number of photons per unit area integrated over the laser pulse length:
\begin{align}
N_{\mathrm{I}} = N_{\mathrm{0}} (1-e^{{-\sigma_{\mathrm{PI}}}\phi}),
\label{fit}
\end{align}

\noindent where  $\sigma_{\mathrm{PI}}$ and the total number of molecules $N_{\mathrm{0}}$ enter as fit parameters. The fit gives an ionization cross section of $\sigma_{\mathrm{PI}}$~=~$\left\langle\sigma(\lambda_{\mathrm{L}})\right\rangle_\mathrm{300K}$~=~(4.7~$\pm$~0.8)~$\times~10^{-16}$~cm$^2$. 

\subparagraph{Data analysis:}
In order to calculate $S_\text{N}$, mass spectra are summed up and subtracted from the background in both measurement modes to obtain $S_\text{res}$ and $S_\text{off}$. There is a systematic error by assuming that the mass signals are proportional to the number of detected molecules. We thus consider a worst case scenario where every event at the detector is attributed to a single detected molecule  $N_\text{event}~=~1$. We then compare the amplitudes of our mass signals within a threshold value, that is derived from the standard deviation of the background noise. The probability of not detecting a particle $P_{\text{zero}}$ is assumed to follow Poissonian statistics
\begin{align}
P_{\text{zero}} = \frac{N_\text{event}}{N_{\text{frames}}} = \text{e}^{-\lambda_\text{P}}.
\end{align}

with $\lambda_\text{P}$ as the average number  of counts per frame. The total number of detected molecules $N$ within one measurement consisting of $N_{\text{frames}}$ is then given by
\begin{align}
N = N_{\text{frames}}(-\ln{P_{\text{zero}}}),
\end{align}

while Gaussian error propagation delivers the 1$\sigma$ error-bars of each data point. 

\subparagraph{Quantum model of the interference fringes:}

Our beam experiments are supported by phase space simulations as introduced by Nimmrichter and Hornberger  for near-field matter-wave interferometry \cite{Nimmrichter2008} and refined for time-domain experiments \cite{Nimmrichter2011a}. 
We here adapt the model to the details of our study. The simulations are based on the one-dimensional Wigner function $w(x,p_x)$ with $x$ and $p_x$ for the positions and the momenta of states, respectively. The Wigner function is defined as the transformation of the position density matrix $\rho(x,x')=\bra{x}\hat{\rho}\ket{x'}$ \cite{Wigner1932}
\begin{equation}
w\left( x,p_x\right)=\frac{1}{2\pi\hbar}\int\, ds\,e^{ip_x s/\hbar}\bra{x-\frac{s}{2}}\hat{\rho}\ket{x+\frac{s}{2}},
\label{Wigner_function}
\end{equation}

\noindent where the molecular beam propagation at a time $t$ is represented by the Hamiltonian $\mathcal{H}_0=p_x^2/2m$ in absence of external fields. The Wigner function therefore transforms like
\begin{equation}
w\left( x,p_x\right) \rightarrow w\left(x-\frac{p_x\, t}{m},p_x\right).
\label{free_propagation}
\end{equation}

\noindent The formalism allows a simple comparison with classical phase space dynamics based on ballistic trajectories. Under free evolution the classical phase space density transforms like the Wigner function \cite{Hornberger2004}. 

Position shifts due to constant accelerations parallel to the grating vectors ($x$-axis) can be included to account for gravitational, electric or magnetic forces. Additionally, a tilt of the molecular beam by the angle $\gamma$  can be introduced as a constant momentum $p_{\gamma}=mv\tan(\gamma)$ parallel to the $x$-axis. The Wigner function for free propagation over a time $t$ therefore reads
\begin{equation}
w\left( x,p_x\right) \rightarrow w\left( x-\frac{p_xt}{m}+\frac{p_{\gamma}t}{m}+\frac{gt^2}{2},p_x-p_{\gamma}-amt\right),
\label{free_propagation_a_v}
\end{equation}

\noindent Transmission through the $k$-th grating $G^{(\text{k})}$ is described by a complex transmission function $t^{(\text{k})}(x)$ acting on the position density matrix
\begin{equation}
\rho(x,x')~\rightarrow~t^{(\text{k})}(x)\rho(x,x')t^{(\text{k})}(x)^*
\label{position_density_transformation}
\end{equation}

\noindent while $\left|t^{(\text{k})}(x)\right|^2$ gives the probability for a particle at position $x$ to remain in the beam and is assumed to follow poissonian statistics
\begin{equation}
\left|t^{(\text{k})}(x)\right|^2=\exp\left(-n^{(\text{k})}(x)\right),
\label{transmission_probability}
\end{equation}

\noindent where $n^{(\text{k})}(x)$ is the number of absorbed photons and shows a $d$-periodic modulation
\begin{equation}
n^{(\text{k})}(x)=n_0^{(\text{k})}\cos^2\left(\frac{\pi x}{d}\right).
\label{photon_absorbtion_number}
\end{equation} 

\noindent While $\left|t^{(\text{k})}(x)\right|^2$ describes a pure absorptive grating, the additional phase modulation is described by
\begin{equation}
\phi^{(\text{k})}(x)=\phi_0^{(\text{k})}\cos^2\left(\frac{\pi x}{d}\right).
\label{phase_modulation}
\end{equation} 

\noindent Here $n_0^{(\text{k})}$ is the average number of absorbed photons in an antinode and $\phi_0^{(\text{k})}$ the eikonal phase, gained by integration of the interaction potential over the intensity profile of the laser. For the optical gratings they read \cite{Rodewald2017}
\begin{equation}
n_0^{(\text{k})}=\frac{4\sigma E^{(\text{k})}\lambda_\text{L}}{hcA_\text{L}}, ~~~~\phi_0^{(\text{k})}=\frac{16\pi^2 E^{(\text{k})}\alpha_\text{L}}{hc A_\text{L}},
\label{phi}
\end{equation} 

\noindent where $E^{(\text{k})}$ is the pulse energy, $A_\text{L}$ the illuminated area, $c$ the speed of light, $\sigma$ the absorption cross section and $\alpha_\text{L}$ the optical polarizability at the grating wavelength $\lambda_\text{L}$. Taking imperfections into account such as a mirror reflectivity $R ~=~0.97$ and a grating coherence factor $C~=~0.76$ \cite{Doerre2015}, there is  an effective reduction of the coherent contribution to $n_0$ 
\begin{equation}
n_{0,\text{eff}}^{(\text{k})} = RC n_0^{(\text{k})}, 
\end{equation}

\noindent while $\phi_{0,\text{eff}}^{(\text{k})}= RC \phi_0^{(\text{k})}$.
Using these parameters the complex transmission function of the optical gratings reads
\begin{equation}
t^{(\text{k})}(x)=\exp\left(-\frac{n_\text{eff}^{(\text{k})}(x)(1+R)}{4 RC}+i\phi_\text{eff}^{(\text{k})}(x)\right).
\label{tansmission_function}
\end{equation} 

\noindent Transmission through a grating is described by the convolution of the Wigner function and the transmission kernel %
\begin{equation}
w\left( x,p_x\right) \rightarrow \int dp_0~T^{(\text{k})}\left(x,p_x-p_0\right)w(x,p_0).
\label{convolution_with_grating}
\end{equation}

\noindent For optical gratings the transmission kernel $T^{(\text{k})}(x,p)$ consists of the Talbot coefficients $B_{\text{n}}^{(\text{k})}(x)$, gained by Fourier expansion of the transmission function $t^{(\text{k})}\left(x\right)$ 
\begin{align}
T^{(\text{k})}\left(x,p_x\right) &=\frac{1}{2\pi\hbar}\sum_n \exp\left(\frac{2\pi inx}{d}\right) \notag \\ &\times \int ds\, e^{ip_x s/\hbar} B_{n}^{(\text{k})}\left(\frac{s}{d}\right),
\label{transmission_kernel}
\end{align}
\begin{widetext}
\begin{align}
B_{n}^{(\text{k})}(\chi) &= \exp\left(\frac{-n_\text{0,eff}^{(\text{k})}}{2}\right)\left(\frac{\sin\left(\pi\chi\right)-\beta\cos\left(\pi\chi\right)}{\sin\left(\pi\chi\right)+\beta\cos\left(\pi\chi\right)}\right)^{\frac{n}{2}}\notag \\
&\times J_n\left(\text{sign}\left(\frac{\sin\left(\pi\chi\right)}{\beta}+\cos\left(\pi\chi\right)\right)\frac{n_\text{0,eff}^{(\text{k})}}{2\beta}\sqrt{\sin^2\left(\pi\chi\right)-\beta^2\cos^2\left(\pi\chi\right)}\right),
\label{Talbot_coefficient}
\end{align}
\end{widetext}

\noindent with the dimensionless parameter $\beta$ as the ratio of molecular absorption cross section and optical polarizability containing information about the electronic structure of the considered molecules
\begin{align}
\beta =\frac{n_0^{(\text{k})}}{2\phi_0^{(\text{k})}}=\frac{\sigma_\text{L}\lambda_\text{L}}{8\pi^2\alpha_\text{L}}.
\label{Beta}
\end{align}

\noindent Transformation (\ref{convolution_with_grating}) also holds for the classical case when we exchange the Talbot coefficients $B_{n}^{(\text{k})}(\chi)$ in (\ref{transmission_kernel}) with  classical coefficients $C_{n}^{(\text{k})}(\chi)$. This can easily be done by substituting  $\sin(\pi\chi)\rightarrow \pi\chi$ and $\cos(\pi\chi)\rightarrow 1$ in (\ref{Talbot_coefficient}) \cite{Nimmrichter2013,Nimmrichter2011}. Note that $B_{n}^{(\text{k})}(\chi)$ is $d$-periodic in $\chi$, whereas $C_{n}^{(\text{k})}(\chi)$ is not. For $\chi = nT/T_\text{T}\rightarrow 0$, $B_{n}^{(\text{k})}(0)=C_{n}^{(\text{k})}(0)$ and describes the behavior of a purely absorptive grating. Therefore the coefficients simplify in terms of the modified Bessel functions $I_{\text{n}}(x)$
\begin{equation}
B_{n}^{(\text{k})}(0)=(-1)^n\,\exp\left(-\frac{n_0^{(\text{k})}}{2}\right)I_{n}\left(\frac{n_0^{(\text{k})}}{2}\right).
\label{Transmission_coeff}
\end{equation}

\noindent This formalism allows to describe the beam propagation through a Talbot-Lau-Interferometer as sequences of free propagation followed by transmission through a grating. The initial state at the first grating is assumed to be an incoherent mixture with a spatial extension $X_0\gg d$ and a momentum $P_0\gg h/d$. The initial Wigner function at the first grating $w_0(x,p_{\text{x}})$ is described by the transverse momentum distribution $D(p_{\text{x}})$, which is gained by integration of the three dimensional momentum density distribution $\mu (p_x,p_{y},p_{z})$ over two dimensions: $D(p_{x})=\int dp_{y}\, dp_{z}~ \mu(p_x,p_y,p_z)$. This leads to
\begin{equation}
w_0\left(x,p_x\right)=\frac{1}{X_0}D(p_x+p_{\gamma}),
\label{w0}
\end{equation}

\noindent where $p_{\gamma}$ denotes the additional constant momentum due to the tilt. According to (\ref{convolution_with_grating}), transmission through the first grating with the transmission kernel $T^{(1)}$ leads to  $w_1\left(x,p_x\right)$. Note that $p_{\gamma}$ is a constant momentum. Therefore the substitution in the integral $\int{dp'T^{(1)}(x,p_x-p')D(p')}$ with $p'=p_0+p_{\gamma}$ leads to $dp'=d(p_0+p_{\gamma})=dp_0$. With iterative usage of (\ref{free_propagation_a_v}) and (\ref{convolution_with_grating}), the Wigner function transforms to $w_2(x,p_x)$ after free propagation for a time $t=T_1$, then to $w_3(x,p_x)$ after transmission through the second grating and after another free propagation for a time $t=T_2$, it transforms to $w_4(x,p_x)$, which denotes the state of the beam before interacting with the third grating. The corresponding transformations are listed below:
\begin{widetext}
\begin{align}
w_1\left( x,p_x\right)&=\frac{1}{X_0}\int dp_0~ T^{(1)}\left(x,p_x-p_0-p_{\gamma}\right)D\left(p_0+p_{\gamma}\right)\notag\\
w_2\left(x,p_x\right)&=\frac{1}{X_0}\int dp_0~T^{(1)}\left(x-\frac{p_x\, T_1}{m}+\frac{p_{\gamma}T_1}{m}+\frac{g\, T_1^2}{2},p_x-p_0-p_{\gamma}-a\, m\, T_1\right)D\left(p_0+p_{\gamma}\right), \notag\\
w_3\left(x,p_x\right)&=\frac{1}{X_0}\int dp_1 T^{(2)}\left(x,p_x-p_1\right)\int dp_0~T^{(1)}\left(x-\frac{p_1\, T_1}{m}+\frac{p_{\gamma} T_1}{m}+\frac{g\, T_1^2}{2},p_1-p_0-p_{\gamma}-a\, m\, T_1\right)D\left(p_0+p_{\gamma}\right),\notag\\\notag 
w_4\left(x,p_x\right)&=\frac{1}{X_0}\int dp_1T^{(2)}\left(x-\frac{p_x\, T_2}{m}+\frac{p_{\gamma} T_2}{m}+\frac{g\,T_2^2}{2},p_x-p_1-p_{\gamma}-a\, m\, T_2\right)\notag\\
&\times\int dp_0\, T^{(1)}\left( x-\frac{p_x\, T_2}{m}-\frac{p_1\, T_1}{m}+\frac{p_{\gamma}\left(T_1+T_2\right)}{m}+\frac{g\left(T_1^2+T_2^2\right)}{2}, p_1-p_0-p_\gamma-a\, m\, T_1\right)D\left(p_0+p_{\gamma}\right). 
\label{trafos}
\end{align}
\end{widetext}

\noindent The third grating masks the fringe pattern of the traversing molecular beam in space. Finally, all molecules are detected independent of their transverse momentum. Therefore only the spatial density distribution of the beam is needed which is calculated by integrating $w_4(x,p_x)$ over the momentum. $\widetilde{D}(x)$ is the Fourier transform of the momentum distribution \cite{Nimmrichter2007,Nimmrichter2013}
\begin{equation}
 \widetilde{D}\left(x\right)=\int dp_x\, e^{-ip_x x/\hbar}D\left(p_x\right).
\label{momentum_distribution}
\end{equation}
\noindent Due to the broad initial momentum, $\widetilde{D}(x)$ is assumed to be very narrow and to peak around $\widetilde{D}(0)=1$. So only index pairs $(k,l)$ which fulfill $\left|kT_1+lT_2\right|\ll T_{\text{T}}$ contribute to the integral of $w_4\left(x,p_x\right)$ in (\ref{trafos}). In the near-resonant and symmetric approximation one assumes $T_1=T$ and $T_2=T+\tau$, where $\tau$ denotes a small delay of the grating timing compared to the Talbot time $\left|\tau\right|\ll T_{\text{T}}$. This restricts the index pairs $(k,l)$ to $k=-l$ and changes the Wigner function $w_4(x)$ to
\begin{align}
w_4\left(x\right)&=\frac{1}{X_0}\sum_{l}\widetilde{D}\left(\frac{l\tau}{T_{\text{T}}}d\right)B_{-l}^{(1)}\left(\frac{l\tau}{T_{\text{T}}}d\right)\notag\\
&\times B_{2l}^{(2)}\left(\frac{l\left(T+\tau\right)}{T_{\text{T}}}\right)\exp\left[\frac{2\pi il}{d}\left(\Delta x\right)\right],
\label{w4_resonance}
\end{align}
\begin{equation}
\Delta x=\Delta x_{\text{s}}-\frac{p_{\gamma}\tau}{m}-gT^2-2g\tau T -\frac{g\tau^2}{2}.
\label{phase_shift}
\end{equation}

\noindent Here $\Delta x_{\text{s}}$ denotes the relative grating shift $\Delta x_{\text{s}} = \Delta x_1 -2\Delta x_2 + \Delta x_3$. The spatial distribution of $w_4(x)$ is scanned using the third grating, which acts as a purely absorptive mask. Therefore one can use $B_{-l}^{(3)}(0)$ from (\ref{Transmission_coeff}). For sufficiently small delays $\tau$ and due to the random phase of the impinging matter wave, $G^{(1)}$ can also be treated as a purely absorptive grating, with $B_{-l}^{(1)}(d\, l\tau /T_{\text{T}})=B_{-l}^{(1)}(0)$. 

Convolution of (\ref{w4_resonance}) with the transmission Kernel $T^{(3)}(x,p)$ and integration over the whole phase space leads to the detected signal $S(\Delta x)$ behind the third grating

\begin{align}
S\left(\Delta x\right)=\sum_lS_{l}~\exp\left[\frac{2\pi il}{d}\Delta x(\Delta x_{\text{s}},T,\tau)\right],
\label{Signal}
\end{align}

\begin{align}
S_{l}=\widetilde{D}\left(\frac{l\tau}{T_{\text{T}}}d\right)B_{-l}^{(1)}\left(0\right)B_{2l}^{(2)}\left(\frac{l\left(T+\tau\right)}{T_{\text{T}}}\right)B_{-l}^{(3)}\left(0\right).
\label{Sl}
\end{align}

\noindent As a proof for interference the periodic modulation of $S\left(\Delta x\right)$ is observed by scanning over the phase of (\ref{Signal}). This can be done either by changing the grating shift  $\Delta x_{\text{s}}$ or the momentum contribution $p_{\gamma} \tau/m$. A slight delay of the third grating timing $\tau$ modulates the phase as well as the signal amplitude, due to the contribution of $\tau$ to the sharp peaked function $\widetilde{D}$.

\begin{figure}[!b]
		\resizebox{1\columnwidth}{!}{\includegraphics{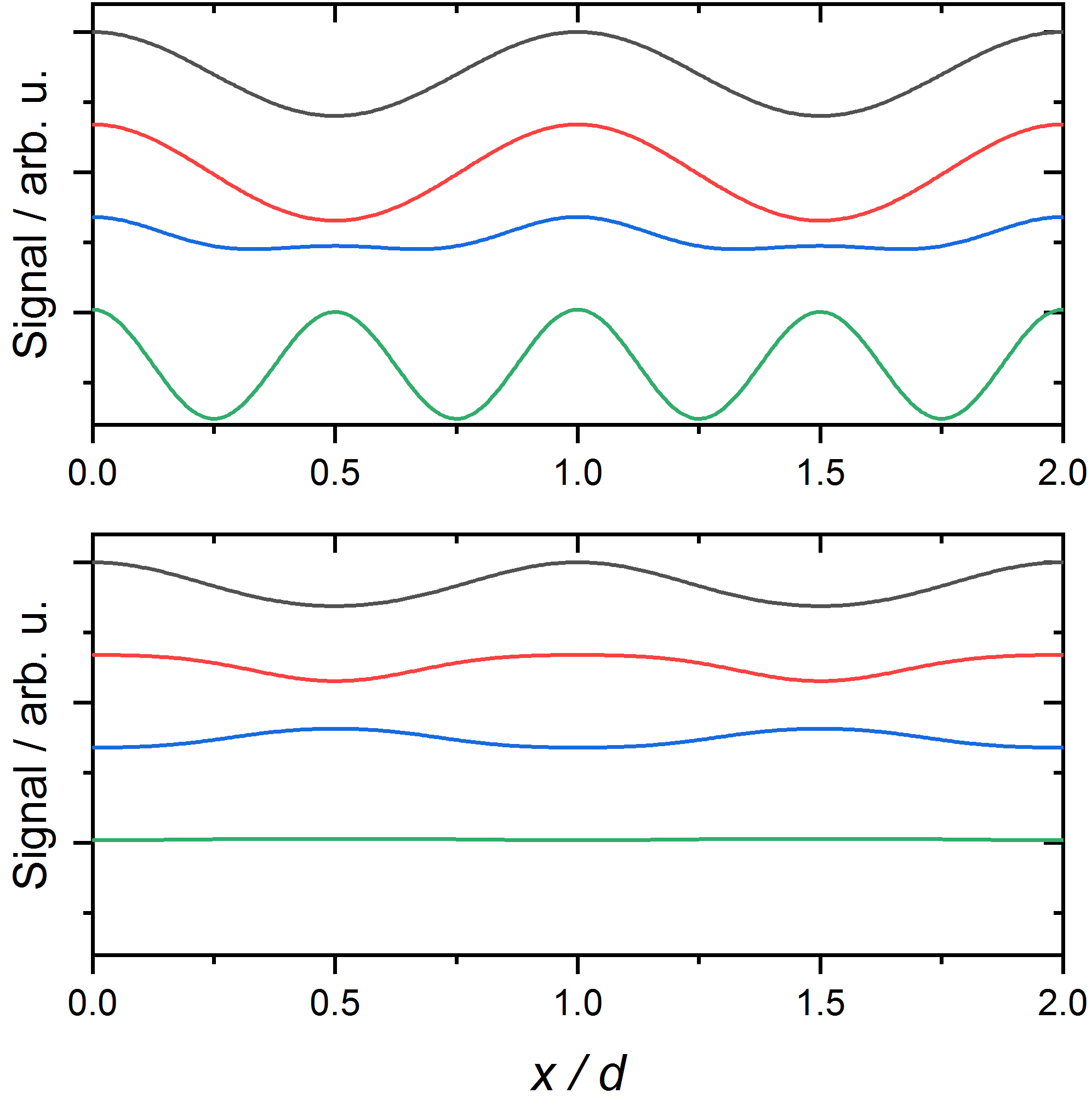}}
		\caption{Signal at $G^{(3)}$ as depending on the grating strength $n_{0,\mathrm{eff}}$ = 3 (black), 4 (red), 6 (blue) and 12 (green) for the $n$~=~1/2 Talbot order for the quantum (top) and the classical (bottom) case, respectively.}
		\label{period} 
\end{figure}

\vspace{0.5cm}
\subparagraph{Periodicity of the observed fringes:}
In 3-grating interferometers with material gratings, observing a shorter fringe period in the fractional order $n$~=~1/2 is a clear proof for quantum interference~\cite{CLAUSER1997121}. Although the simulations in Figure~\ref{Dip} already show that quantum interference has been observed by comparing the measured normalized signal to simulations based on a phase space model, it might appear unintuitive that the fringes in the fractional order show the same periodicity as in the first Talbot order. Figure~\ref{period} shows the expected quantum or classical signal at $G^{(3)}$ for the fractional Talbot order at several grating strengths, i.e. different mean number of absorbed photons $n_{0,\mathrm{eff}}$. For the grating strengths used in our experiments $n_{0,\mathrm{eff}}$~=~3 in both the $n$~=~1 and the $n$~=~1/2 Talbot orders $d$-periodic fringes are expected.

\end{document}